\documentclass{aa}  

\usepackage{graphicx}
\usepackage{txfonts}
\usepackage{amsmath}
\usepackage{tabularx}
\usepackage{longtable}
\usepackage{lineno}
\usepackage{hyperref}
\usepackage{xcolor}
\usepackage{braket}
\usepackage{multirow}
\usepackage{subfig}
\usepackage[utf8]{inputenc}
\hypersetup{
colorlinks=false, 
linkcolor=blue,
urlcolor=red, 
linktoc=all 
}
\bibpunct{(}{)}{;}{a}{}{,} 

\usepackage{subcaption}
\usepackage{comment}

\begin{document} 

\title{Frequency-dependent time delays in Sgr~A* from a broadband VLA observation}

\author{Noemi La Bella
          \inst{1}
          \and Heino Falcke
            \inst{1}
          \and Christiaan Brinkerink
            \inst{1}
          \and Jesse Vos
            \inst{2}
          }

\institute{
Department of Astrophysics, Institute for Mathematics, Astrophysics and Particle Physics (IMAPP), 
Radboud University, P.O. Box 9010, 6500 GL Nijmegen, The Netherlands\\
\email{n.labella@astro.ru.nl}
\and
Centre for mathematical Plasma Astrophysics, Department of Mathematics, KU Leuven, Celestijnenlaan 200B, B-3001 Leuven, Belgium
}

\abstract
{Frequency-dependent radio time delays have been measured in Sgr~A* light curves, with higher-frequency emission generally leading lower-frequency emission. These frequency-dependent lags have been interpreted in terms of an organized jet-like outflow, expanding plasma components, or magnetic-flux eruptions, although their origin and recurrence remain uncertain.}
{We aim to search for and quantify frequency-dependent radio time delays in a new high-cadence Karl G. Jansky Very Large Array (VLA) observation of Sgr~A*. Using the improved broad 12--37~GHz frequency coverage on the VLA, we measured the timing of the dominant variability feature and compared the resulting lag--wavelength slope with previous measurements.}
{Sgr~A* was observed with the VLA on 28 May 2022 in the Ku, K, and Ka bands over an $\sim6$~hour track. We extracted both band-averaged and 2~GHz subband light curves and measured the delays using a peak-time method and a locally normalized cross-correlation function.}
{The light curves show one broad variability feature across the observed frequency range, with the maximum occurring progressively later at lower observing frequencies. Both timing methods recover a positive lag--wavelength relation, with a slope of order $40~\mathrm{min\,cm^{-1}}$. This value agrees with time lag results reported over more than 15 years from the first VLA observations at $22$--$43$~GHz to later multifrequency VLA and ALMA campaigns.}
{ The sign and magnitude of the estimated delay are consistent with outward propagation through an optically thick synchrotron-emitting region. The recurrence of comparable delays over almost two decades suggests that frequency-dependent radio lags are a persistent feature of Sgr~A* variability. Broadband, multi-epoch VLA monitoring of Sgr~A* provides a useful probe of its radio variability, but robust lag measurements require high cadence and improved $(u,v)$ coverage of both Sgr~A* and the surrounding extended emission such as by combining different VLA configurations.}

\keywords{Galaxy: center -- black hole physics -- radio continuum: general -- techniques: interferometric -- accretion, accretion disks}

\maketitle
\nolinenumbers

\section{Introduction}
\label{sec:intro}

Sagittarius~A* (Sgr~A*) is the compact radio source associated with the supermassive black hole at the center of the Milky Way. Its mass, $M\simeq4\times10^{6}\,M_{\odot}$, is constrained by stellar orbital measurements \citep{Ghez2008,Gillessen2009,Gravity2020}. Event Horizon Telescope (EHT) observations  at $\sim230$~GHz have resolved the emission on horizon scales, revealing a ring-like morphology with a central brightness depression \citep{EHT2022I,EHT2022II,EHT2022III,EHT2022IV,EHT2022V,EHT2022VI}, interpreted as the black hole shadow \citep{Falcke2000}. At radio to submillimeter wavelengths, the spectral energy distribution of Sgr~A* rises toward a peak around 1--2~THz, consistent with self-absorbed synchrotron emission \citep{Falcke1998}. The source is optically thick at lower radio frequencies and becomes increasingly optically thin at higher frequencies, so multifrequency observations probe different depths in the compact emitting region. Jet models for Sgr~A* reproduce its flat radio spectrum as a superposition of self-absorbed synchrotron components \citep{Blandford1979,Falcke1993}. Based on this model, \citet[][Sec. 3.3.]{Falcke1999} predicted frequency-dependent time lags and sizes as a signature of a jet-like outflow. 
Very long baseline interferometry (VLBI) measurements also show that the apparent size decreases with the observed wavelength once interstellar scattering is accounted for, indicating that higher frequencies sample smaller spatial scales closer to the black hole \citep{Bower2006,Doeleman2008,Issaoun2019}.

Sgr~A* is variable throughout the electromagnetic spectrum, with short infrared and X-ray flares and lower-amplitude radio and millimeter variability on timescales from minutes to hours \citep{Dexter2014,Witzel2018,Wielgus2022}. GRAVITY resolved loop-like centroid motion and polarization-angle rotation during bright near-infrared flares, consistent with compact emission evolving on dynamical timescales close to the black hole \citep{GRAVITY2018,GRAVITY2023}. At millimeter wavelengths, ALMA observations show variability on timescales of tens of minutes to hours \citep{Wielgus2022}.
In addition to the characteristic variability timescale, time delays have also been measured between light curves observed at different frequencies. \citet{YusefZadeh2006} reported that the peak of a flare at 43~GHz preceded that at 22~GHz by 20--40~min. They explained the delay using the expanding cloud model of \citet{vanDerLaan1966}, in which an initially optically thick synchrotron-emitting component expands and first becomes optically thin at higher frequencies and subsequently at lower frequencies.  \citet{Falcke2009} interpreted frequency-dependent size and lag relations in terms of the jet model and derived a moderately relativistic outflow velocity with $\gamma \beta \sim2$.
More recently, general relativistic magnetohydrodynamics (GRMHD) simulations by \citet{Vos2024} showed that outward-propagating magnetic-flux eruptions in the jet-disk boundary can produce lower-frequency delays through the combined effects of plasma propagation and evolving synchrotron opacity. Using simultaneous Very Large Array (VLA) and ALMA observations between 19 and 100~GHz, \citet{Brinkerink2015} measured a lag--wavelength slope of $42\pm14~\mathrm{min\,cm^{-1}}$. Assuming that the variability traces an outflow, the estimated lag and intrinsic size--frequency relation implied mildly relativistic flow speeds. In a subsequent multi-epoch VLA study between 18 and 48~GHz, \citet{Brinkerink2021} find positive lag slopes in five of eight epochs. Their synthetic data analysis was consistent with a persistent lag relation of approximately $40~\mathrm{min\,cm^{-1}}$, which they interpreted as evidence of a mildly relativistic outflow.
At lower frequencies, \citet{Michail2021} measured a delay of approximately 20~min between 9.948 and 8.052~GHz within a single VLA receiver band and epoch. At millimeter wavelengths, \citet{miyazaki2013} estimated a much shorter lag of -2.56~min between 102 and 90~GHz, with the lower-frequency light curve leading the higher-frequency one.
This change in lag behavior at higher frequencies may be expected as the emission approaches the transition from the optically thick to the optically thin synchrotron regime \citep{Falcke2009,Brinkerink2015}.  At even higher frequencies, \citet{Wielgus2022} report no measurable lag between the $213$ and $229$~GHz ALMA light curves; however, the fractional frequency separation $\Delta\nu/\nu$ between these bands is small, and the resulting sensitivity may not be sufficient to rule out detectable time delays across the broader submillimeter range. Finally, \citet{Brinkerink2021} argue that short-timescale variability correlated across several frequency bands can introduce a zero-lag component in cross-correlation curves, while uncorrelated noise can bias the recovered lags toward smaller values.

We present VLA observations of Sgr~A* obtained on 28 May 2022, spanning $12$--$37$~GHz with high temporal resolution. To our knowledge, no dedicated VLA multifrequency lag study has been published since \citet{Brinkerink2021}, whose multi-epoch sample covered observations taken in 2015.
The primary aim of this work is to measure the frequency-dependent timing of the dominant radio variability feature across the VLA bands and to compare the resulting lag--wavelength slope with previous measurements. 
Earlier VLA lag programs interleaved regular scans on an external phase calibrator with the on-source scans. Our observations were instead designed to exploit the bright minispiral emission in Sgr~A West as an intra-field amplitude reference, testing for the first time at VLA frequencies the calibration approach developed for ALMA by \citet{Mus2022} and \citet{Albentosa-Ruiz2025}. This strategy provides continuous on-source coverage of Sgr~A*, enabling higher-cadence light curves for the lag analysis.

The paper is organized as follows. 
Section~\ref{sec:reduction} describes the observations, the standard data reduction procedure, and the intra-field minispiral calibration. In Sect.~\ref{sec:results}, we present the light curves across the VLA frequency bands and sub-bands and report the time lag analysis. Section~\ref{sec:discussion} discusses the results obtained compared with previous works. The conclusions are presented in Sect.~\ref{sec:conclusions}.

\section{Data reduction}
\label{sec:reduction}
\subsection{A priori calibration}
Sgr~A* was observed with the VLA in A configuration, with a maximum baseline of approximately 36~km, on 28 May and 12 June 2022. The latter epoch is not included in the present analysis because the calibration did not provide a reliable flux-density scale across the bands.  On 28 May, Sgr~A* was observed between 05:44:28 and 11:52:26~UTC, for a total duration of approximately $6$~hours. The schedule cycled through the three receiver bands in sequence with $\sim1$~min scans per band, yielding a nominal on-source cadence of $\sim3$~min per band. Requantizer setup scans of $\sim30$~s between each frequency-tuning change were flagged, doubling the effective per-band cadence to $\sim4.5$~min. The only interruptions of the Sgr~A* coverage were a few breaks of $\sim9$~min for pointing solutions on NRAO~530.
The observations covered the Ku ($\sim 12$--$18$~GHz), K ($\sim 18$--$26$~GHz), and Ka ($\sim 29$--$37$~GHz) receiver 
bands, divided into spectral windows ($48$ for Ku, $64$ for K and Ka) of $128$~MHz bandwidth and $64$ channels of $2$~MHz each. After flagging, the lag analysis used effective frequencies between $\sim13$ and $\sim36$~GHz. The K band was strongly affected by the water-vapor absorption line at $ \sim22$~GHz and was only partially considered in the analysis of the light curve. 
The data were recorded in full polarization using the $3$-bit samplers and the WIDAR correlator. The present work uses only the parallel-hand correlations; the polarization products will be considered in a future analysis. The data were calibrated in \texttt{CASA~v6.7}. Initial flagging removed zero-valued data, radio-frequency interference, and the first second of each scan. 
Gain-curve, antenna-position, and atmospheric-opacity corrections 
were applied with CASA \texttt{gencal}, using the standard VLA reference tables for the gain-elevation curve and the seasonal opacity model at the VLA site. The absolute flux-density scale, together with instrumental delays, bandpass responses, and initial complex-gain solutions, was derived from 3C~286 with the \citet{Perley2017} model, and pointing corrections were obtained from NRAO~530. Because no separate phase-calibrator cycle was included, phase corrections were obtained through self-calibration on Sgr~A* itself, using projected baselines longer than $150~\mathrm{k}\lambda$, where Sgr~A* is effectively unresolved and the surrounding Sgr~A West emission was filtered out. The visibility phase on these baselines is therefore intrinsically close to zero, so any phase deviation can be associated with instrumental and atmospheric contributions. To correct residual time-dependent amplitude variations, we then tested an intra-field amplitude calibration based on the minispiral emission, as discussed in the next section.

\begin{figure*}
    \centering
    \includegraphics[width=\linewidth]{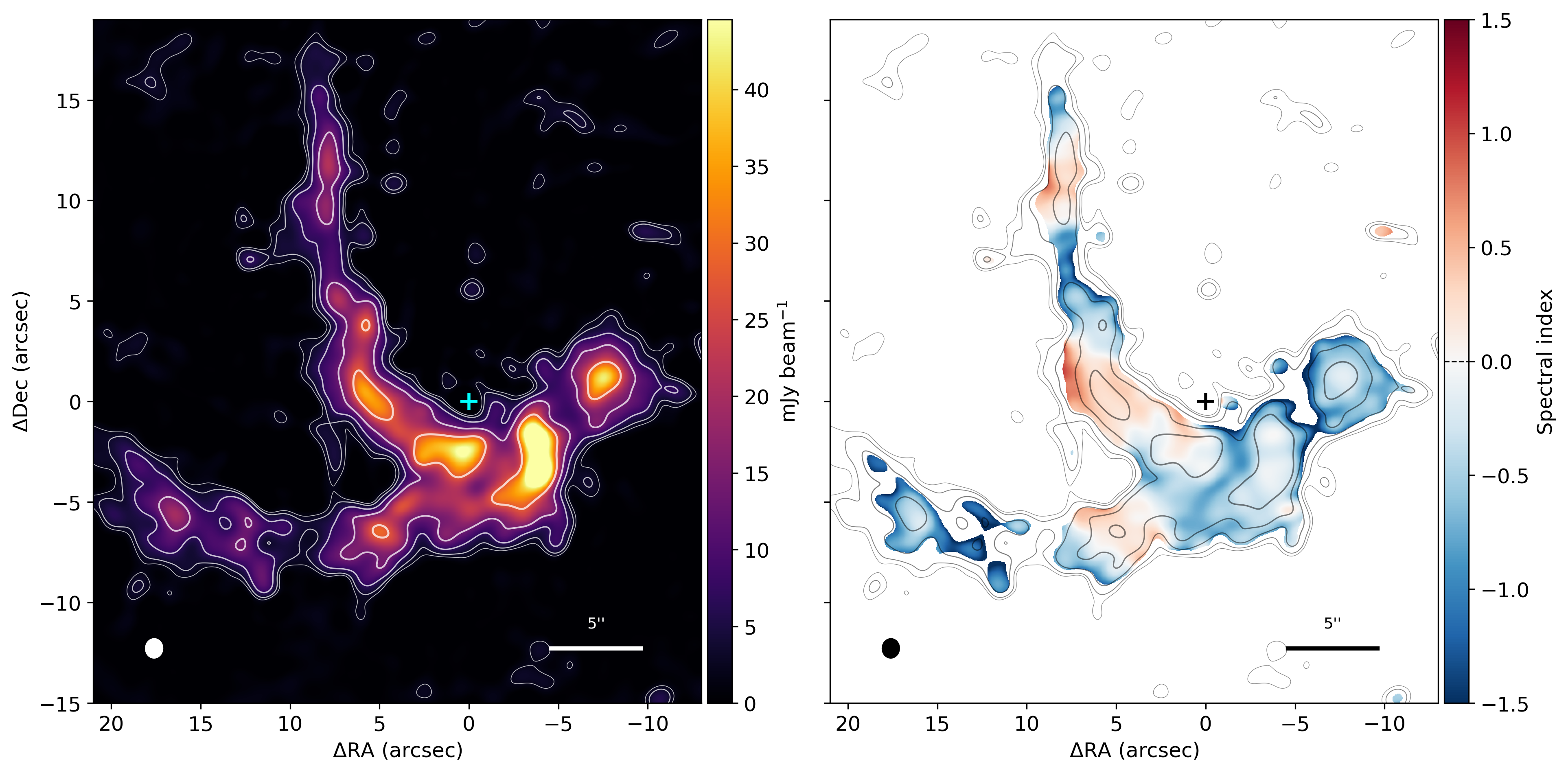}
        \caption{Left: Multiband radio continuum image of  Sgr~A West obtained by combining Ku, K, and Ka bands over a 6-hour observation. The Ku, K, and Ka images were constructed using baselines shorter than 150$~\mathrm{k}\lambda$ and convolved to a common Ku-band beam of $0.96 \times 0.95$~arcsec before combination.  The position of Sgr~A* is indicated by the cross.  Contours show $3, 5, 10, 20, 40, 80 \times \sigma_{\rm rms}$, where $\sigma_{\rm rms}$ is the noise level measured in an off-source region.  Right:  Spectral index map $\alpha$ ($S_\nu \propto \nu^\alpha$), derived  from the three band images at  $\sim15$, $\sim19$, and $\sim32 \,\mathrm{GHz}$. Only pixels detected above $5\sigma$ in all three bands are shown. Intensity contours from the left panel are overlaid.}
 
    \label{fig:minispiral}
\end{figure*}

\subsection{Minispiral calibration test} 
\label{sec:minispiral}
Sgr~A West, also known as the ``minispiral,'' is a complex ionized-gas structure surrounding Sgr~A* extending over angular scales of several tens of arc seconds \citep[e.g.,][]{Zhao2003}. Because this emission is bright and is not expected to vary over the duration of a single observation, it can be used as an intra-field amplitude calibrator.
\citet{Mus2022} and \citet{Albentosa-Ruiz2025} developed an intra-field calibration method for ALMA observations of Sgr~A*, which uses the extended emission from Sgr~A West as amplitude reference. The method constructs a two-component visibility model comprising the extended minispiral emission and the compact source, fits it to the visibilities at each solution interval, and uses the fitted minispiral scale factor to correct residual time-dependent amplitude variations. For more details, we refer to \citet[][]{Albentosa-Ruiz2025}.

We adapted this approach to the VLA data. 
To construct a minispiral-only CLEAN model, we first estimated the compact Sgr~A* flux density for each scan and spectral window from the mean real part of the calibrated visibilities on baselines $>150~\mathrm{k}\lambda$, where the extended minispiral is largely resolved out. This value was subtracted from the data for the corresponding scan and spectral window. The residual short-baseline visibilities were imaged, and the Fourier transform of this CLEAN model gives the extended-emission visibility model, $V^{\rm ext}$.
We then fitted the original calibrated visibilities, restricted to projected baselines $<150~\mathrm{k}\lambda$, with a two-component model, $V(t) = S_1(t)\,V^{\rm ext} + S_2(t)\,V^{\rm compact}$, where $V^{\rm compact}$ is the visibility of a point source at the phase center. The scale factors $S_1(t)$ and $S_2(t)$, representing the minispiral and compact-source components, respectively, were obtained by weighted least squares at each solution interval. A successful amplitude calibration should leave $S_1(t)$ approximately constant, with the intrinsic compact-source variability appearing in $S_2(t)$.

Several minispiral model images were tested: models built from averaged 2~GHz sub-bands, intermediate spectral window groupings, and the full band, as well as two image cell sizes ($0.03$ and $0.25\,\mathrm{arcsec}$). Solution intervals of one scan of $\sim$1~min, 30~min, and 60~min were tested for each configuration. We also repeated the fits using longer and shorter short-baseline cuts. In all cases, $S_1(t)$ exhibited strong variation of $\sim30-50 \%$, with no clear systematic trend. This instability of $S_1(t)$ exceeds the level expected from instrumental or atmospheric gain, as verified by inspecting the amplitude gain solutions on 3C~286 and NRAO~530, and most likely reflects the changing $(u,v)$ sampling of the asymmetric extended structure.

The minispiral covers approximately the central $80$~arcsec around Sgr~A*, with internal substructure on arc-second scales \citep{Zhao2009,Kunneriath2012}. The corresponding spatial frequencies range from a few thousand to several tens of thousands of wavelengths. The VLA was in its most extended configuration (A) during our observations, and most baselines are insensitive to the minispiral structure, with only a small number of antennas contributing to the relevant short baselines in each scan (e.g., $3$--$6$ antennas in the highest-frequency band). As the array rotates over the 6-hour track, different regions of the asymmetric minispiral visibility function are sampled per scan, and a single per-scan scaling $S_1(t)$ cannot capture this structural variation. Increasing the solution interval to $30$ and $60$~min averages over a wider $(u,v)$ range but does not remove the $S_1(t)$ instability. The minispiral amplitude corrections were consequently not applied to the final dataset, and the reported flux densities rely on the a priori amplitude calibration from 3C~286 combined with the phase self-calibration on Sgr~A*.

Figure~\ref{fig:minispiral} shows the multiband radio continuum image of Sgr~A West, obtained by combining the Ku, K, and Ka band images over the 6-hour observation. This image is intended to highlight the extended minispiral emission rather than the highest-resolution compact source. For this reason, the three band images were constructed using only visibilities on baselines shorter than 150$~\mathrm{k}\lambda$, which removes the long-baseline information that provides the highest angular resolution. Sgr~A* was subtracted before imaging, and its position is indicated by a cross. 
The spectral index $\alpha$, defined as $S_\nu \propto \nu^\alpha$, was computed pixel by pixel from the three band-averaged images, after convolution to a common beam and reprojection onto a common grid. The three main components of Sgr~A West are clearly detected: the northern arm, the eastern arm, and the bar \citep[e.g.,][]{Ekers1983,Paumard2004,Nitschai2020}. The spectral index distribution shows a mixture of positive and slightly negative values across the minispiral, with regions of the bar and northern arm exhibiting $\alpha \approx - 0.1$, {with typical per-pixel 1$\sigma$ uncertainties of $\sim$ 0.04 ,  consistent with optically thin free-free bremsstrahlung emission from ionized gas. The overall picture is qualitatively consistent with the multifrequency spectral index studies of the minispiral \citep[e.g.,][]{Kunneriath2012}. Values of $\alpha$ significantly more negative than $- 0.1$ may reflect residual calibration uncertainties or the fact that extended emission is resolved out more at higher frequencies, so the Ka band recovers less flux on large scales than Ku.

\begin{figure}[h]
\centering
\includegraphics[width=\linewidth]{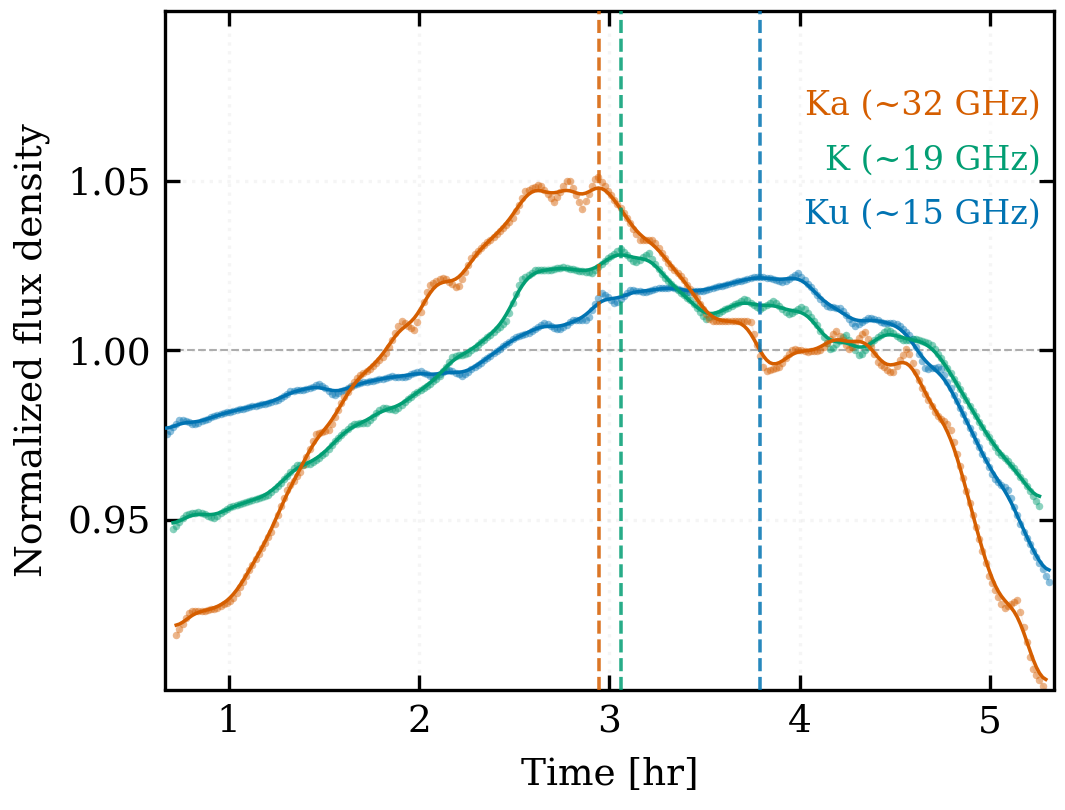}
\caption{
Band-averaged Sgr~A* light curves for the 28 May 2022 observation. 
The curves show the Ku, K, and Ka bands, normalized by their median flux density over the analyzed time window. 
The solid curves show the smoothed light curves, and the dashed vertical lines mark the corresponding peak times. 
}
    
\label{fig:lightcurves_bands}
\end{figure}

\begin{figure}[h]
\centering
\includegraphics[width=\linewidth]{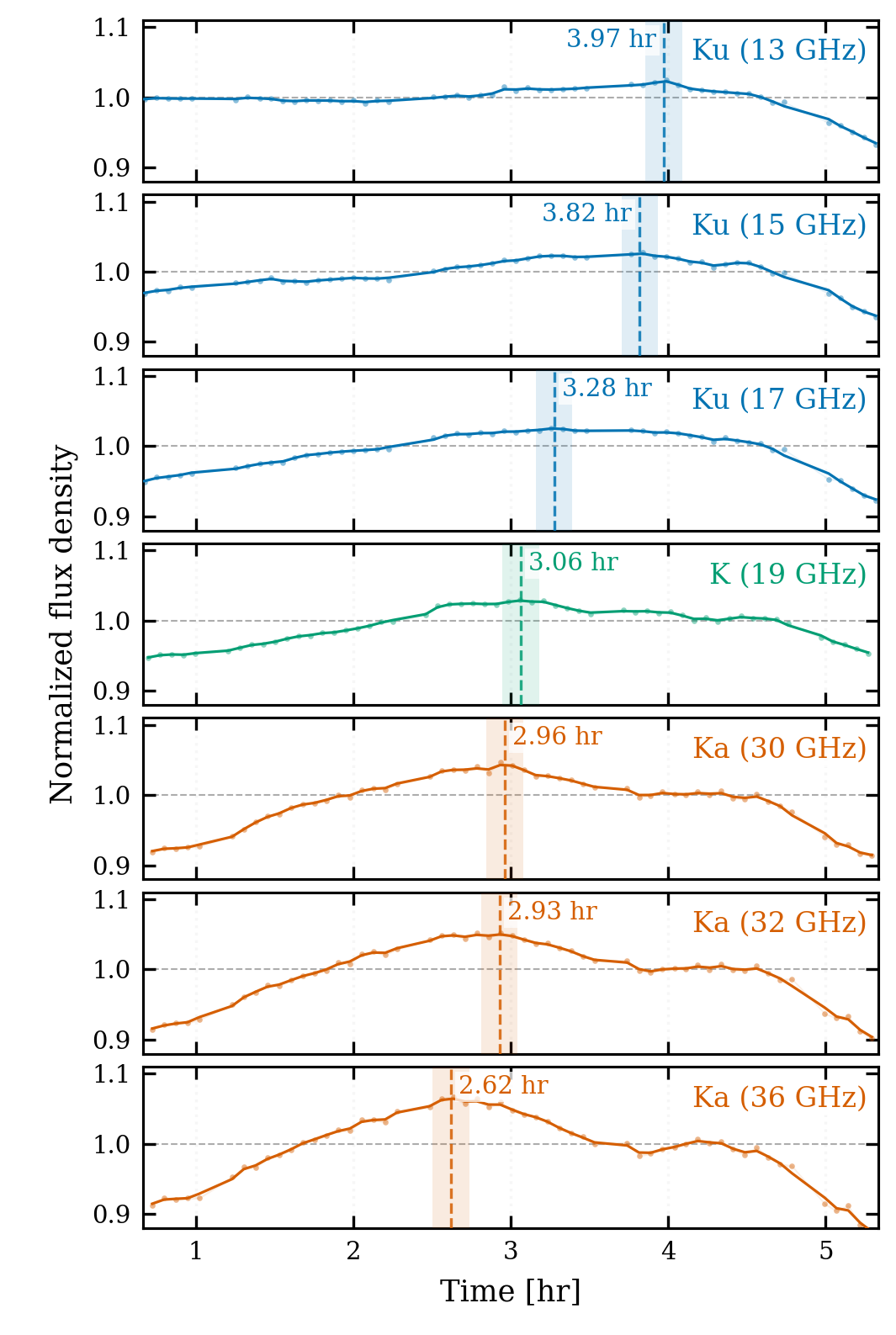}
\caption{Sub-band Sgr~A* light curves for the 28 May 2022 observation, ordered from lowest to highest observing frequencies. Each light curve is normalized by its median flux density over the analyzed time window. The solid curves show the smoothed light curves, the dashed vertical lines mark the measured peak times, and the shaded $\pm 7$ min regions indicate the visual peak-time windows. Frequency labels are rounded to the nearest gigahertz for readability.}
\label{fig:lightcurves_subbands} 
\end{figure}

\section{Results}
The Sgr~A* light curves were extracted directly from the calibrated visibilities after a priori calibration and phase self-calibration. In the following section, we present the frequency-resolved radio light curves and use them to measure time delays as a function of observing frequency.
\label{sec:results}

\subsection{Light curves extraction}

To reduce contamination from Sgr~A West, we extracted the light curves on projected baselines longer than $150  \ {\rm k}\lambda$ \citep{Brinkerink2015} where Sgr~A* is dominant. For each scan, Stokes~$I$ was formed from the two parallel-hand correlations. Since Sgr~A* was placed at the phase center, the flux density was estimated from the real part of the selected visibilities. The uncertainty at each point was estimated from both the visibility weights and the scatter of the real visibility values within the scan, and the larger of the two was adopted. A $2$-$3\%$ fractional systematic uncertainty from the 3C286 amplitude calibration \citep{Perley2017} was added in quadrature. Scans with fewer than ten unflagged visibilities after the baseline selection were discarded.

We considered two extraction strategies. In the first, the selected visibilities were averaged directly within each 2~GHz sub-band after applying the baseline cut, yielding one flux-density measurement per scan. In the second, the flux density was first extracted independently for each spectral window and the resulting spectral-window light curves were then combined using inverse-variance weighting. Both approaches gave consistent broad light-curve morphologies and peak ordering. We adopted spectral-window-based extraction for the analysis below, because it preserves frequency information within each receiver band and allows us to test for intra-band delays.

Because no stable intra-field amplitude correction was applied, residual time-dependent amplitude errors may remain in the light curves. These can arise from residual gain-transfer errors, atmospheric coherence losses, pointing, or opacity effects, especially at low elevation. To further reduce them, we excluded the first and the final $40$~min of the observation corresponding to the lowest elevations of the track ($\lesssim 20^\circ$), where phase-coherence losses become significant, as discussed by \citet{Brinkerink2015}.

Figure~\ref{fig:lightcurves_bands} shows the band-averaged Ku, K, and Ka light curves (~$\sim15$ GHz, $\sim19$ GHz and $\sim32$ GHz, respectively), while Fig.~\ref{fig:lightcurves_subbands} shows the 2~GHz light curves of each sub-band, centered at $\sim13$, $\sim15$, and $\sim17$~GHz in the Ku band; $\sim19$~GHz in K band; and $\sim30$, $\sim32$, and $36$~GHz in Ka band. These effective frequencies were obtained after flagging and inverse-variance weighting. An additional Ka sub-band centered near $\sim34$~GHz was inspected but excluded from the analysis because it was significantly noisier and did not provide a robust light curve.

The curves were normalized by their median flux density over the analyzed time window and averaged in $1$~min bins, comparable to the scan duration. The binned data are smoothed with a Gaussian kernel of full width at half maximum (FWHM) of $7$~min. The dashed vertical lines mark the peak times measured with the procedure described in Sect.~\ref{sec:timelags}.
The three receiver-band light curves show a similar broad rise, maximum, and decline. The maximum occurs first at the Ka band, followed by the K band and then the Ku band, consistent with the well-known frequency-ordered sequence reported for VLA observations of Sgr~A* radio variability \citep{YusefZadeh2006,Brinkerink2015,Brinkerink2021}. The same trend and frequency ordering is also visible in the 2~GHz sub-band light curves in Fig.~\ref{fig:lightcurves_subbands}: the peak shifts from earlier to later as the observing frequency decreases. The quantitative measurements of these inter-band and intra-band delays are reported in Table~\ref{tab:lags} and discussed in Sect.~\ref{sec:timelags}. A weaker secondary feature is visible near $ \simeq 4$~hours, most clearly in the highest-frequency Ka curve. Since this feature is not equally prominent in all frequency bands, we do not interpret it as a separate intrinsic component and do not use it for the lag measurement.

Furthermore, the peak amplitude appears larger at higher frequencies, as found in previous radio light curves of Sgr~A* \citep[e.g.,][]{Falcke1998,Herrnstein2004,YusefZadeh2006,Brinkerink2021}. As a consistency check, the mean flux densities recovered on baselines longer than $150~\mathrm{k}\lambda$ are approximately $0.65$, $0.78$, and $0.93$~Jy in the Ku, K, and Ka bands, respectively. These values correspond to a spectral index of $\alpha = 0.45 \pm 0.05$, which is consistent within uncertainties with the compact-source spectral indices $\alpha = 0.41 \pm 0.03$ (VLA+ALMA) and $\alpha = 0.50 \pm 0.07$ (VLA-only) reported by \citet{Brinkerink2015}.
\subsection{Time lags}
\label{sec:timelags}

The sub-band light curves show a similar broad evolution of flux density across the analyzed time window: a slow rise, a broad maximum, and a subsequent decline. As noted in the light curve plots, the timing of this broad maximum appears to shift to later times at lower observing frequency. We therefore measured the frequency-dependent time delays using two complementary methods. First, we used a non-parametric peak-time estimator, which measures the time of the broad maximum in each light curve. Second, we computed the locally normalized cross-correlation function \citep[LCCF;][]{Welsh1999}, which measures the relative delay from the full rise-and-decline morphology of the light curves. The two approaches have previously been used in studies of Sgr~A* radio lags \citep[e.g.,][]{Brinkerink2015,Brinkerink2021,Vos2024}. Throughout this section, positive lags mean that the lower-frequency light curve is delayed relative to the reference light curve.

\subsubsection{Peak-time method}
The peak-time method measures the relative timing of the dominant broad variability feature, following the same idea as the curve-fitting approach used by \citet{Brinkerink2015}. However, we did not impose a fast-rise exponential decay (FRED) profile. The feature in our light curves is broad and roughly symmetric around the maximum, so a parametric fast-rise exponential-decay model is not needed, and we used a nonparametric peak estimator instead. Each sub-band light curve was normalized by its median flux density, binned in $1$~min intervals, and smoothed with a Gaussian kernel of FWHM $7$~min. The smoothing suppresses short-timescale scatter while preserving the overall shape and timing of the broad feature. The peak time was taken as the maximum of the smoothed curve and refined to sub-bin precision using a quadratic fit through the three points around the discrete maximum. Short-timescale structure can shift the recovered peak position, so we estimated the peak-time uncertainty using a Monte Carlo method: in each realization we randomly kept $50\%$ of the data points, reapplied the same peak estimator, and repeated the procedure $500$ times. The standard deviation of the recovered peak times is the peak-time uncertainty.

The highest-frequency Ka sub-band reaches the broad maximum first and was used as the reference. Relative to this reference, the lowest-frequency Ku sub-band is delayed by $81.1 \pm 12.5$~min. The individual sub-band lags increase systematically with wavelength, so the broad feature peaks later at lower observing frequency.

Figure~\ref{fig:timelagsfit} shows the peak-time lags as a function of wavelength. The reference sub-band defines zero lag and is not treated as an independent measurement in the fit. To quantify the wavelength dependence, we fitted a linear relation defined as
\begin{equation}
\Delta t(\lambda) = \mathcal{S}_{\rm peak}\,(\lambda - \lambda_{\rm ref}) + c,
\label{Eq:fitting}
\end{equation}
where $\Delta t(\lambda)$ is the peak-time lag at wavelength $\lambda$ relative to the $35.93$~GHz reference, indicated as $\lambda_{\rm ref}$, $\mathcal{S}_{\rm peak}$ is the lag--wavelength slope, and $c$ is a constant intercept. The fit uses inverse-variance weights from the peak-time uncertainties and gives

\begin{equation}
\mathcal{S}_{\rm peak} = 41.2 \pm 10.2~\mathrm{min\,cm^{-1}}.
\end{equation}
For robustness, we performed several checks. Fitting the receiver-band averaged lags gives $\mathcal{S}_{\rm peak} = 43.5 \pm 9.1~\mathrm{min\,cm^{-1}}$. 
Including the reference point in the fit gives $\mathcal{S}_{\rm peak} = 46.7 \pm 8.0~\mathrm{min\,cm^{-1}}$. Changing the reference to the lowest-frequency Ku sub-band at $\sim13$~GHz yields $42.3 \pm 10.8~\mathrm{min\,cm^{-1}}$. Finally, using a Gaussian smoothing kernel of FWHM $5$~min instead of $7$~min gives 
$40.9 \pm 10.6~\mathrm{min\,cm^{-1}}$. All tests are consistent within the uncertainties.

\begin{figure}[h]
\centering
\includegraphics[width=\linewidth]{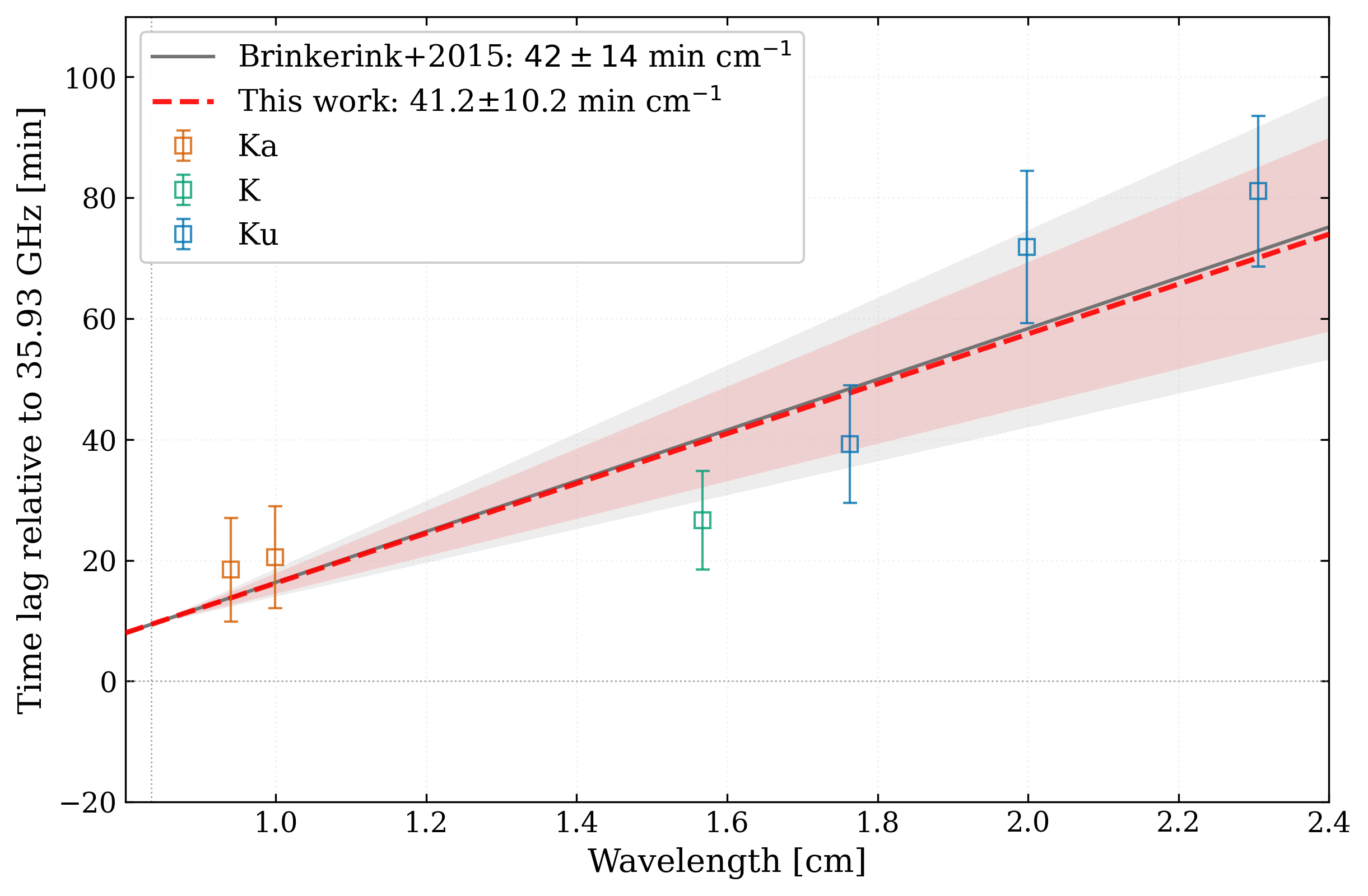}
\caption{
Peak-time lag as a function of observing wavelength, measured relative to the 35.93 GHz Ka reference sub-band. The points display the individual sub-band lag measurements, colored by receiver band. The dashed red line and shaded region show the weighted fit to the individual sub-band lags and its ($1\sigma$) uncertainty. The gray line and shaded region show the lag--wavelength slope measured by \citet{Brinkerink2015}, shifted to the same reference convention for comparison.
}

\label{fig:timelagsfit}
\end{figure}

\begin{figure*}[h!]
\centering
\includegraphics[width=\linewidth]{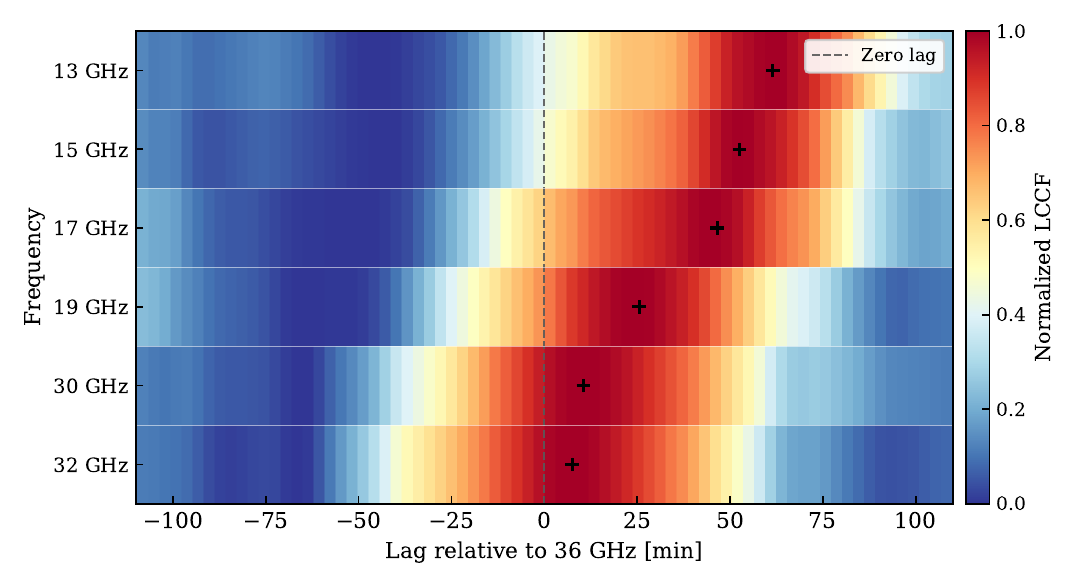}
\caption{
Normalized LCCF as a function of time lag for each sub-band, computed relative to the highest-frequency Ka reference sub-band at 35.93~GHz. Positive lags indicate that the lower-frequency light curve is delayed relative to the reference. The vertical dashed line marks zero lag, and the crosses mark the recovered LCCF maxima. The frequency labels are rounded to the nearest gigahertz for readability. The exact effective frequencies and lag values are listed in Table~\ref{tab:lags}.
}
\label{fig:heatmap}
\end{figure*}
\subsubsection{Locally normalized cross-correlation function}

As an independent check, we computed the LCCF \citep{Welsh1999} between the same sub-band light curves used in the peak-time analysis, with the method outlined in \citet{Vos2024, Brinkerink2021}. For each light curve, the LCCF was calculated against the $35.93$~GHz Ka reference. 

A direct cross-correlation of the full light curves can be affected by short-timescale variations that are present in several sub-bands near zero lag. \citet{Brinkerink2015} already pointed out that this structure can dominate the cross-correlation near zero lag and mask the delayed evolution of the broader feature. We therefore restricted the LCCF analysis to the central part of the observation, from $100$ to $290$~min, which contains the rise and decline of the broad feature and excludes the noisier edges of the light curves. We estimated the uncertainty on each LCCF lag with an analogous Monte Carlo resampling of the peak-time method, in which $50\%$ of the points in both light curves were randomly retained, the LCCF was recomputed, and the lag of its peak was remeasured.

Figure~\ref{fig:heatmap} shows the normalized LCCF as a function of time lag for each sub-band. The vertical dashed line marks zero lag, and the crosses mark the recovered LCCF maxima. The Ka sub-bands have small positive lags, while the K and Ku sub-bands show increasingly delayed correlation peaks. The largest delay is found for the lowest-frequency Ku sub-band, with an LCCF lag of $58.4 \pm 12.1$~min relative to the $35.93$~GHz reference. The LCCF maxima therefore move systematically to positive lag at lower observing frequency.

We fitted the individual LCCF lags with the same linear relation as the peak-time analysis (Equation~\ref{Eq:fitting}). This gives

\begin{equation}
\mathcal{S}_{\rm LCCF} = 40.4 \pm 5.8~\mathrm{min\,cm^{-1}}.
\end{equation}
As for the peak-time method, we repeated the fit with different fitting conventions and alternative reference sub-bands. The recovered slopes are consistent within the uncertainties.
The peak-time and LCCF analyses therefore agree on both the sign and the magnitude of the lag--wavelength relation. The two slopes are consistent with previous radio measurements of Sgr~A* \citep{YusefZadeh2006,Brinkerink2015,Brinkerink2021,Michail2021}, as discussed in the next section. A summary of peak times and lags is listed in Table~\ref{tab:lags}.

\begin{table}[h]
\centering
\setlength{\tabcolsep}{3pt}
\caption{Peak times and time lags of the  Sgr~A* sub-band light curves.}
\label{tab:lags}
\begin{tabular}{lcccc}
\hline\hline
Band & Frequency & Peak time & Lag (peak) & Lag (LCCF) \\
     & (GHz)     & (min)     & (min)      & (min)      \\
\hline
Ka & $35.93$ & $157.3 \pm 5.7$  & $0$ (ref.)        & $0$ (ref.)        \\
Ka & $31.87$ & $175.7 \pm 6.3$  & $+18.4 \pm 8.5$   & $+7.2  \pm 3.3$   \\
Ka & $29.99$ & $177.8 \pm 6.2$  & $+20.5 \pm 8.4$   & $+10.5 \pm 5.2$   \\
K  & $19.12$ & $183.9 \pm 5.8$  & $+26.6 \pm 8.1$   & $+22.5 \pm 13.0$  \\
Ku & $17.00$ & $196.5 \pm 7.9$  & $+39.2 \pm 9.8$   & $+43.5 \pm 10.1$  \\
Ku & $15.00$ & $229.2 \pm 11.2$ & $+71.9 \pm 12.6$  & $+53.1 \pm 7.9$   \\
Ku & $13.00$ & $238.4 \pm 11.1$ & $+81.1 \pm 12.5$  & $+58.4 \pm 12.1$  \\
\hline
\multicolumn{3}{l}{Slope $\mathcal{S}$ ($\mathrm{min\,cm^{-1}}$)} 
                 & $41.2 \pm 10.2$ & $40.4 \pm 5.8$ \\
\hline
\end{tabular}
\tablefoot{Frequencies are the inverse-variance-weighted mean frequency of each $2$~GHz sub-band after flagging. The peak times and corresponding peak-time lags are measured within the $40$--$320$~min window. The LCCF lags are measured within the restricted $100$--$290$~min window. The uncertainties are estimated from Monte Carlo resampling. All lags are measured relative to the $35.93$~GHz reference sub-band. The slope $\mathcal{S}$ in the last row is the weighted linear fit to the sub-band lags via Eq.~\ref{Eq:fitting}, excluding the reference point.
}
\end{table}

\section{Discussion}
\label{sec:discussion}

The peak-time method and LCCF analyses both find that the higher-frequency light curves lead the lower-frequency ones. The two methods measure the delay in different ways: the peak-time method uses the timing of the broad maximum, whereas the LCCF uses the lag that most closely aligns the rise-and-decline morphology of the light curves. As a result, the individual sub-band lags are not identical. In particular, the peak-time delay between the highest-frequency Ka reference and the lowest-frequency Ku sub-band is $81.1 \pm 12.5$~min, compared with $58.4 \pm 12.1$~min from the LCCF. The smaller absolute LCCF lags may reflect the restricted time window used for the analysis ($100$--$290$~min). This window excludes the noisier observation edges and isolates the central part of the broad variability feature, where the cross-correlation is most reliable.

Despite these differences, the recovered lag--wavelength slopes agree within the uncertainties: $\mathcal{S}_{\rm peak}=41.2 \pm 10.2~\mathrm{min\,cm^{-1}}$ and $\mathcal{S}_{\rm LCCF}=40.4 \pm 5.8~\mathrm{min\,cm^{-1}}$. The consistency of the recovered slopes suggests that the measured lag--wavelength relation is not driven by the choice of timing method. We therefore used the lag--wavelength slope as the main quantity for comparison with previous works.

Our result agrees with \citet{Brinkerink2015}, who reported a lag--wavelength slope of $42\pm14~\mathrm{min\,cm^{-1}}$ from simultaneous ALMA and VLA observations in 2012. \citet{Brinkerink2021} later analyzed eight VLA epochs from 2015 and reported positive lag--wavelength relations in five of them. Combining the time lag values from the three epochs with the largest lag--wavelength slopes, which are close in value, they obtain a fitted slope of $40.3\pm2.0~\mathrm{min\,cm^{-1}}$. They suggest that the difference between detections and non-detections may reflect either intrinsic source variability or differences in data quality. Our 2022 epoch provides an independent measurement seven years after the most recent observation in their sample, with a clear broad feature present in the Ku, K, and Ka light curves and a lag--wavelength slope in the same range. For direct comparison, Fig.~\ref{fig:timelagsfit} shows the \citet{Brinkerink2015} slope shifted to the same reference wavelength as our fit.

The similar sign and magnitude of these measurements have been interpreted as indicating outward motion in the radio-emitting region. By combining their measured lags with the Sgr~A* size--frequency relation and inclination constraints \citep{Bower2006}, \citet{Brinkerink2015} and \citet{Brinkerink2021} inferred mildly relativistic outflow speeds ($\gamma\beta \sim0.5-1.2$), consistent with the compact-jet interpretation discussed by \citet{FalckeMarkoff2000} and \citet{Falcke2009}. Our measured slope agrees with this scenario. However, the lag--wavelength relation alone does not uniquely distinguish a jet-like outflow from an adiabatically expanding synchrotron component \citep{vanDerLaan1966}. The outflow interpretation relies on comparing the measured lags with the VLBI size--frequency relation of Sgr~A* \citep{Bower2006}. This comparison assumes that the variable component producing the lags follows the same size--frequency relation as the bulk radio emission \citep{Falcke2009}, but a distinct expanding component could have a different size--frequency evolution. Nevertheless, no separate variable subcomponents have been identified so far in VLBI observations of Sgr~A*, so the jet-like outflow interpretation is still a plausible explanation.

More recently, \citet{Vos2024} demonstrated that detailed radiative post-processing of GRMHD simulations exhibiting prominent magnetic-flux eruptions in the jet-disk boundary naturally reproduces positive lag--wavelength relations with the observed sign and order of magnitude. The steepest slopes, reaching $\sim40~\mathrm{min\,cm^{-1}}$, occur when emission from erupting flux tubes contributes substantially to the total flux, as in epochs with clear flaring activity such as those discussed by \citet{Brinkerink2021} and in the present work.
Although their simulations were performed at higher frequencies than the 13--36~GHz band considered here, which precludes a direct light-curve comparison, they show that the dynamic interplay among multiple emission components, together with their emissivities and absorptivities, naturally reproduces the diversity of observed lag--wavelength relations.  
Their simulations yield a broad distribution of slopes, with a concentration in the $0$--$40~\mathrm{min\,cm^{-1}}$ range. This suggests that the intrinsic variability of the lag--wavelength relation may be considerably greater than is currently inferred from observations, while the comparatively narrow observational range may also reflect limited sampling, selection effects, and other observational biases. However, the predicted radio lags depend partly on the adopted particle-acceleration prescription and electron distribution. These assumptions affect the centimeter-radio emissivity, optical depth, and model spectrum and may also influence the comparison between the predicted and measured lags.

The intra-band lags provide a qualitative consistency check. Within Ku band, the $13$ and $15$~GHz sub-bands lag the $17$~GHz sub-band by $\sim42$ and $\sim33$~min, respectively. These lags are broadly consistent with the positive intra-band delay reported by \citet{Michail2021} between $8.05$ and $9.95$~GHz. In Ka, the $32$ and $30$~GHz sub-bands also lag the $35.93$~GHz reference by $\sim18$ and $\sim20$~min. The Ka intra-band delays are expected to be smaller than the Ku ones because the wavelength range within the band is shorter, and the per-sub-band peak times are less precise due to worse atmospheric coherence at higher frequencies. The K band cannot contribute to the intra-band check because flagging around the 22~GHz water absorption line yields only one usable $2$~GHz sub-band at $\sim19$~GHz. The intra-band delays therefore show the same sign of the global lag--wavelength trend, but are less constraining than the full inter-band fit.

Two limitations should be considered when interpreting this result. First, the analysis is based on a single epoch, so it cannot by itself determine how often this lag--wavelength relation occurs. Second, no minispiral amplitude correction was applied to the final light curves, and some residual per-scan amplitude structure may remain, particularly in the highest-frequency band. Such residuals could arise from elevation-dependent gain variations, atmospheric-opacity errors, coherence losses, or imperfect amplitude transfer from the calibrators. To reduce these effects, we excluded the first and final $40$~min of the observation, when the source was at the lowest elevations of the track, below $\sim20^\circ$. Nevertheless, the lag measurement depends primarily on the relative timing of the broad variability feature rather than on the absolute flux-density scale.  Furthermore, residual amplitude errors are not expected to produce the observed frequency ordering, with the peak appearing first at the highest frequencies and progressively later at lower frequencies. The frequency-ordered peak times, together with the agreement between the smoothed-peak and LCCF analyses, suggest that the recovered slope is not dominated by residual amplitude calibration effects.

\section{Conclusions}
\label{sec:conclusions}
We presented $\sim6$-hour VLA observations of Sgr~A* obtained on 28 May 2022 in the Ku, K, and Ka bands, using a receiver setup nominally covering $12$--$37$~GHz and providing an effective cadence of about $4.5$~min per band. The aim was to search for frequency-dependent time delays in the radio variability of Sgr~A* to compare the resulting lag--wavelength relation with previous VLA measurements and interpret the possible origin of the frequency-dependent lags.

The light curves show one broad variability feature across the observed frequency range, with the maximum occurring first at the highest frequencies and progressively later at lower frequencies. We recover a positive lag--wavelength relation with a slope of order $40~\mathrm{min\,cm^{-1}}$, comparable to the values reported by \citet{Brinkerink2015} from the combined 2012 ALMA and VLA observations and by \citet{Brinkerink2021} from the 2015 VLA campaign. Together with the earlier detections reported by \citet{YusefZadeh2006,YusefZadeh2008}, our 2022 measurement extends the evidence for positive radio time lags in Sgr~A* over a time span of nearly two decades. The sign and magnitude of the delay are consistent with radio variability moving outward through an optically thick synchrotron-emitting region, where shorter-wavelength emission is seen before longer-wavelength emission, as expected in compact-jet and outflow models \citep{FalckeMarkoff2000,Falcke2009}. 
The recurrence of comparable positive frequency-dependent delays is notable because changes in the accretion flow of Sgr~A* may occur on decade timescales. For instance, wind-capture simulations predict that the mass accretion rate, and consequently the emission from the inner accretion region, can vary by $\lesssim20$--$40\%$ on timescales of $<100$~yr, even though the mass flux from the stellar sources remains constant \citep{Melia2001}. The passage of G2 was also expected in some models to enhance accretion and Sgr~A* activity \citep[e.g.,][]{Schartmann2012}, yet monitoring during the encounter yields no evidence for a change in the mean flux density or spectrum of Sgr~A* \citep{Bower2015}. The repeated detection of similar positive lag trends, together with VLBI evidence that different radio wavelengths probe different spatial scales in Sgr~A* \citep{Bower2006,Falcke2009}, suggests that its frequency-dependent radio structure remains broadly stable over long timescales, in spite of short-term variability during individual flares.

Finally, we explored the application of an intra-field amplitude calibration scheme to VLA observations of Sgr~A*, following the approach of \citet{Mus2022} and \citet{Albentosa-Ruiz2025}.  The method could not be applied to the final lag analysis because the extended A configuration provides sparse $(u,v)$ coverage at the minispiral spatial scales, resulting in unstable scale factors. Future observations combining an extended VLA configuration with a more compact one (for example, A+D) would provide better $(u,v)$ coverage of both Sgr~A* and the extended minispiral, potentially making this approach more effective for high-cadence monitoring of Sgr~A*.

\begin{acknowledgements}
This work is supported by the ERC Synergy Grant ``Blackholistic'' and is part of the project Dutch Black Hole Consortium (with project number 1292.19.202) of the research program NWA, which is (partly) financed by the Dutch Research Council (NWO).
J.V. acknowledges support from Research Foundation -- Flanders (FWO) postdoctoral fellowship 1255226N and KU Leuven postdoctoral mandate PDMT1/24/012.
J.V. acknowledges support from the Dutch Research Council (NWO) and SURF which provided computing resources in support of this work through grants No. NWO-2024.004 and No. NWO-2024.041.
\end{acknowledgements}

\bibliographystyle{aa.bst}
\bibliography{biblio.bib} 

\end{document}